\documentclass[twoside,twocolumn]{article}

\usepackage{blindtext} % Package to generate dummy text throughout this template 

\usepackage{graphicx}
\usepackage{subfigure}

\usepackage[sc]{mathpazo} % Use the Palatino font
\usepackage[T1]{fontenc} % Use 8-bit encoding that has 256 glyphs
\usepackage{microtype} % Slightly tweak font spacing for aesthetics

\usepackage[english]{babel} % Language hyphenation and typographical rules

\usepackage[hmarginratio=1:1,top=32mm,columnsep=20pt]{geometry} % Document margins
\usepackage[hang, small,labelfont=bf,up,textfont=it,up]{caption} % Custom captions under/above floats in tables or figures
\usepackage{booktabs} % Horizontal rules in tables

\usepackage{lettrine} % The lettrine is the first enlarged letter at the beginning of the text

\usepackage{enumitem} % Customized lists
\setlist[itemize]{noitemsep} % Make itemize lists more compact

\usepackage{abstract} % Allows abstract customization
\usepackage{titlesec} % Allows customization of titles
\renewcommand\thesection{\Roman{section}} % Roman numerals for the sections
\renewcommand\thesubsection{\roman{subsection}} % roman numerals for subsections
\titleformat{\section}[block]{\large\scshape\centering}{\thesection.}{1em}{} % Change the look of the section titles
\titleformat{\subsection}[block]{\large}{\thesubsection.}{1em}{} % Change the look of the section titles

\usepackage{fancyhdr} % Headers and footers
\usepackage{titling} % Customizing the title section

\usepackage{hyperref} % For hyperlinks in the PDF

\newcommand{\ra}{\rangle}
\newcommand{\la}{\langle}

\newcommand{\beq}{\begin{equation}}
\newcommand{\eeq}{\end{equation}}

\newcommand{\ball}{\begin{align}}
\newcommand{\eall}{\end{align}}

\newcommand{\beqar}{\begin{eqnarray}}
\newcommand{\eeqar}{\end{eqnarray}}

\newcommand{\ben}{\begin{enumerate}}
\newcommand{\een}{\end{enumerate}}

\pretitle{\begin{center}\huge\bfseries} % Article title formatting
\posttitle{\end{center}} % Article title closing formatting
\title{Comment on arXive:1807.08572: a plausible explanation of the giant diamagnetism found in Au-Ag nanostructures} % Article title
\author{%
\textsc{Navinder Singh}\thanks{Cell Phone: +919662680605} \\[1ex] % Your name
\normalsize Physical Research Laboratory, Ahmedabad, India. \\ % Your institution
\normalsize \href{mailto:navinder.phy@gmail.com}{navinder.phy@gmail.com} % Your email address
}
\date{\today} % Leave empty to omit a date
\renewcommand{\maketitlehookd}{%
\begin{abstract}
\noindent In a recent comment (arXive:1906.05742) on the preprint (arxive:1807.08572) entitled "Coexistence of Diamagnetism and Vanishingly Small Electrical Resistance at Ambient Temperature and Pressure in Nanostructures", it is pointed out that the reduction of the four-probe resistance ($R_{4P}$) to zero value is accompanied by a rise in the two probe resistance ($R_{2P}$) in the same temperature range. This curious correlation between $R_{4P}$ and $R_{2P}$ is said to be pointing towards a  non-superconducting "conductance percolation" transition (rather than "percolating superconducting transition" as proposed in the revised version of arxive:1807.08572). The explanation offered in preprint arXive: 1906.05742 is quite reasonable, but the author leaves open the question of  giant diamagnetism. In this short comment I suggest a plausible cause of the giant diamagnetism found in nanostructures. It could be due to gapped electronic energy spectrum of  nanoparticles which is due to quantum confinement effects, and that  suppresses  the electronic scattering mechanism leading to a very high value of Langevin diamagnetism.

\end{abstract}
}

\begin{document}

% Print the title
\maketitle

%----------------------------------------------------------------------------------------
%	ARTICLE CONTENTS
%----------------------------------------------------------------------------------------

In the revised version, arxive:1807.08572, entitled "Coexistence of Diamagnetism and Vanishingly Small Electrical Resistance at Ambient Temperature and Pressure in Nanostructures", and another preprint, arXive: 1906.02291, entitled "Current-Voltage characteristics in Ag/Au nanostructures at resistive transitions",  Thapa, Pandey, and collaborators presents detailed information about their measurements of zero resistance and diamagnetic transitions found in some of the nanostructered samples prepared by them. It is concluded that in some of the nanostructures exhibit superconducting-like transition at ambient conditions.

In a recent comment (arXive: 1906.05742) it is suggested that the transition observed by IISc team could be a percolating conductance threshold rather than a percolating superconducting transition.  This is nicely explained using figure 2 in arXive: 1906.05742 and it is argued that minute internal structural deformations due to thermal stress (on changing temperature)  lead to some sort of re-distribution of current channels in the nanostructure which can further lead to four probe terminals having same potential. In other words no potential difference between them! This could be quite possible and a possible signature of it is the simultaneous rise in two probe resistance when four probe resistance drops to zero (refer to figure 2 in arXive: 1906.05742). 

In this communication, I point out that the giant diamagnetic transition observed in the nanostructures does not necessarily  imply superconducting transition.  My main point is the following. Induced Langevin diamagnetic susceptibility when an external magnetic field is applied is given by

\beq
\chi_L = -N^* \frac{e^2 \la r^2\ra}{6 m}.
\eeq

This is generally small as $\sqrt{\la r^2\ra}$ is of the order of atomic radius. Now, imagine if $\sqrt{\la r^2\ra}$ is stretched to the  radius of a nanoparticle $\sqrt{\la R^2\ra}$!  Atomic radius is in Angstrom range where as nanoparticle radius is in nanometers.  Thus there could be an order of magnitude larger effect! But there are two complications: (1) there must be some mechanism which suppress the electronic scattering of electrons that constitute surface currents, and (2) we need to estimate the number of such {\it long surviving} electrons at the surface of a nanoparticle. This number of surface electrons should be sizable to enhance the susceptibility (that factor $N^*$ in equation (1)). One mechanism is the well known superconducting transition in which surface currents along the periphery of nanoparticles survive to very long time (via supercurrents). But there is another very important mechanism of suppression of electronic scattering at the surface as pointed out in\cite{1}. It is something to do with gapped electronic states on the surface of nanoparticles. If this gap $\Delta$ is greater than $k_B T$, thermal scattering is suppressed, and surface currents can survive longer which further can lead to giant diamagnetism through equation (1),  with $\la r^2\ra$ replaced with $\la R^2\ra$ where $R$ is the radius of the nanoparticle.  In the following we first calculate discrete electronic structure on the surface of a nanoparticle with a simple minded calculation, and then calculate the fraction of surface electrons. Next we calculate a critical radius of a nanoparticle by comparing the electronic energy gap with thermal energy. Particles with smaller radius than this critical radius can show giant diamagnetism. Then we calculate the number of surface electrons for which thermal scattering is suppressed and advance arguments why electronic scattering is suppressed when there is a gap in the excitation spectrum. Next, we will calculate susceptibility and compare that with atomic susceptibility and show the possibility of giant diamagnetism. We end by discussing some important factors that are left out in these back-of-the-envelop calculations.

\begin{figure}[!h]
\begin{center}
\includegraphics[height=2.7cm,width=5cm]{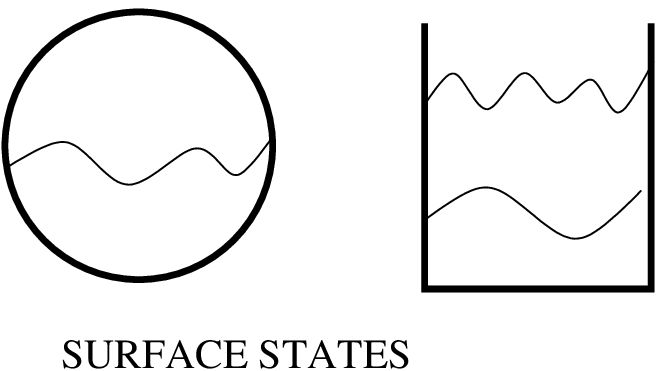}
\caption{Discrete electronic states on the surface of a nanoparticle.}
\end{center}
\end{figure}
Let $R$ be the radius of a nanoparticle. Imposing periodic boundry condition on a closed orbit of a surface elecron along the peripheri of the nanoparticle (figure 1), we get quatized wavevectors $n\frac{2\pi}{L}$. The $n$th energy level is given by $E_n = \frac{n^2 \hbar^2}{2 m R^2}$, and the energy gap by
\beq
\Delta E_n = E_{n+1} -E_n = \frac{\hbar^2}{2 m R^2} (2 n+1).
\eeq
When this gap becomes bigger that $k_B T$ for sufficiently larger $n$ electronic scattering will be suppressed. But before working that out we need to calculate how many levels are populated at the surface, and this requires an estimate of the number of surface electrons. The number of surface electrons are estimated in the following way:

Let $a$ be the lattice constant for the nanoparticle. And assume that area $a^2$  contributes one electron at the surface. Then the area $4\pi R^2$ contains $\frac{4 \pi R^2}{a^2}$ electrons. In the interior of the nanoparticle, assume that volume $a^3$ contributes one electron. Then the volume of nanoparticle will contain $\frac{4 \pi R^3}{3 a^3}$ electrons. The ratio $\alpha$ of the number of surface electrons to the number of volume electrons is given by

\beq
\alpha = 3 \frac{a}{R.} 
\eeq
 
If $a = 1 ~A$ and $R= 1~nm$, then $\alpha = 0.3$.  That if there are $N=10,000$ electrons in the interior of the nanoparticle, then 3000 will reside on the surface (there are several complicating factors which are discussed at the end of this note). A small fraction of those will be able to carry long lived surface currents. To calculate that number let us first calculate the number of filled surface energy levels ($n$). Clearly $2 n = \alpha N$ where factor of 2 is for spin degeneracy. As $n>>1$, the energy gap (equation 2) for the highest filled surface levels is given by

\beq
\Delta E_n^* \simeq \frac{\alpha \hbar^2 N}{2 m R_*^2}.
\eeq

For the suppressing of scattering we must have $\Delta E_n^* > k_B T$. Thus, there is an upper limit to the radius at a given temperature

\beq
R^*(T) \leq \sqrt{\frac{3 \hbar^2 N}{2 mK_B T}\left(\frac{a}{R}\right)}.
\eeq

This is quite an important result but subject to some corrections (as explained in the last paragraph). If radius of a nanoparticle is less that $R^*(T)$, then it can support long lived surface currents and can show giant diamagnetism. Particles with bigger radius will not show diamagnetism.\footnote{This should be tested experimentally.} 

Next, we address the question of the suppression of the scattering rate for a gapped system. It was shown in\cite{2} that if there is a gap in the electronic density of states around the Fermi level in a metal, the electronic scattering is exponentially suppressed:

\beq
\frac{1}{\tau(T)} \simeq e^{-\frac{\Delta}{k_B T}} f(T/\Theta_D).
\eeq

This expression was obtained using the memory function formalism\cite{3}. $f$ is a scaling function and $\Theta_D$ is the Debye temperature\cite{2}. It should not be difficult to repeat the calculation for discrete energy levels of a nanoparticles, and an expression for the scattering rate can be obtained. We assume that an expression similar to the above will be obtained which will show suppressed scattering rate. This is quite obvious, it will only change the details not the main argument (refer also to\cite{1}).

For the calculation of diamagnetic susceptibility, we need to calculate the number of surface electrons that can support long lived surface currents. It is calculated in the following way:
 
The number of surface electrons for which thermal energy is of the order of the electronic energy gap is given by
\beq
N_{cut} = \frac{2 m R^2 k_BT}{\hbar^2}.
\eeq 
These electrons will be thermally scattered and cannot participate effectively in diamagnetizing surface currents. The electrons in which we are interested are those which are thermally NOT scattered, and there number $N^*$ is given by: (total number of surface electrons ($\alpha N$)) - ($N_{cut}$).

\beq
N^* = \alpha N -  \frac{2 m R^2 k_BT}{\hbar^2}.
\eeq

Collecting all this, we finally come to our  main result: Diamagnetic susceptibility of the nanoparticle:

\beq
\chi^{nano}_L = - \left(\alpha N -\frac{2 m R^2k_BT}{\hbar^2}\right) \frac{e^2 R^2}{6 m}.
\eeq

To get a feel for the enhanced effect let us compare it with atomic susceptibility ($\chi^{atomic}_L = N\frac{e^2 a^2}{6 m}$) originating from the interior of the nanoparticle:

\beq
\frac{\chi_L^{nano}}{\chi_L^{atomic}} = \left(\frac{R}{a}\right)^2 \left(\alpha -\frac{2 mR^2k_BT}{N \hbar^2}\right).
\eeq

The second factor containing $K_B T$ is an order of magnitude small as compared to $\alpha$ for the relevant set of parameters. Thus it can be neglected. Substiting the value of $\alpha$ we get

\beq
\frac{\chi_L^{nano}}{\chi_L^{atomic}} \sim 3  \left(\frac{R}{a}\right) \sim 30.
\eeq

This is an order of magnitude larger effect! This could be a cause behind the giant diamagnetic effect seen in IISc experiments on nanostructures. There are several important factors that we did not take into account in the above simple minded calculations. We collect those in the following paragraph (without going into their detailed calculations). 

{\it Left out important factors}:

Although Fermi energies of bulk Silver and bulk Gold are the same ($\sim 5.5~eV$), but for a nano-scaled Silver particle it is modified due to quantum confinement effects. When two metals of different Fermi energies are joined together diffusion of electrons from higher Fermi energy metal to lower Fermi energy metal happens. Thus, there must be some diffusion of electrons from bulk Gold to Silver nanoparticle, and it will renormalize the number of surface electrons on Ag nanoparticle. The factor $\alpha$ has to be multiplied with another factor $f_{Diffusion}$ to take into account the diffusion of electrons. There is another complicating factor. Quantum confinement is more severe in the interior of the nanoparticle (for linear dimension $2R$) than on the surface (linear dimension $2\pi R$). Thus a slight fraction of electrons would like to move to the surface, renormalizing $\alpha$ to $\alpha f_{conf}$ where $f_{conf}$ is a factor that takes into account this quantum confinement effects. There is another factor which is the Coulomb factor. Diffusion to the surface can lead to slight charge imbalance, and there will be electron-electron interactions. This will further renormalize $\alpha$ to $\alpha f_{Coul}$. Collecting all this $\alpha$ will be normalized to $\alpha f_{Diffusion} f_{conf} f_{Coul}$  (A detailed investigation is needed for the computation of these factors. Here we will not go into that).  This will change the ratio of susceptibilities to
\beq
\frac{\chi_L^{nano}}{\chi_L^{atomic}} \sim 3 f_{Diffusion} f_{conf} f_{Coul} \left(\frac{R}{a}\right),
\eeq
leading to some further enhancement of the effect. Thus, in conclusion, there is a possibility of giant diamagnetism without invoking superconductivity!

%----------------------------------------------------------------------------------------
%	REFERENCE LIST
%----------------------------------------------------------------------------------------

%----------------------------------------------------------------------------------------


\begin{thebibliography}{99} % Bibliography - this is intentionally simple in this template


\bibitem{1}A. Hernando, A. Ayuela, P. Crespo, and P. M. Echenique, New J. Phys. {\bf 16}, 073043 (2014).

\bibitem{2}Pankaj Bhalla and Navinder Singh, Eur. Phys. J. B {\bf 89}, 40 (2016).

\bibitem{3}Navinder Singh, {\it Electronic Transport Theories: From Weakly to Strongly Correlated Materials}, CRC Press (2016). 


 
\end{thebibliography}
\end{document}